\documentclass[a4paper,fleqn,usenatbib]{mnras}

\usepackage{newtxtext,newtxmath}
 
\usepackage[T1]{fontenc}
\usepackage{ae,aecompl}

\usepackage{graphicx}	
\usepackage{amsmath}	
\usepackage{amssymb}	
\usepackage{color}
\usepackage{diagbox}

\newcommand{\dnu}{$\Delta\nu$}
\newcommand{\numax}{$\nu_{\rm max}$}
\newcommand{\echelle}{$\acute{e}$chelle}
\newcommand{\lz}{$l=0$}
\newcommand{\lo}{$l=1$}
\newcommand{\lt}{$l=2$}
\newcommand{\kepler}{\emph{Kepler}}
\newcommand{\tess}{\emph{TESS}}

\title[Mode Classification]{Mode Angular Degree Identification in Subgiant Stars with Convolutional Neural Networks based on Power Spectrum }

\author[Minghao Du et al.]{
Minghao Du$^{1}$\thanks{E-mail:du.minghao@mail.bnu.edu.cn},
Shaolan Bi$^{1}$\thanks{\,E-mail:bisl@bnu.edu.cn},
Xianfei Zhang$^{1}$,
Yaguang Li$^{2,3,1}$,
Tanda Li$^{4,3}$ and
Ruijie Shi$^{1}$
\\
$^{1}$Department of Astronomy, Beijing Normal University, Beijing 100875, China\\
$^{2}$Sydney Institute for Astronomy (SIfA), School of Physics, University of Sydney, NSW 2006, Australia\\
$^{3}$Stellar Astrophysics Centre, Department of Physics and Astronomy, Aarhus University, Ny Munkegade 120, DK-8000 Aarhus C, Denmark\\
$^{4}$School of Physics and Astronomy, University of Birmingham, Birmingham, B15 2TT, United Kingdom\\
}

\date{Accepted XXX. Received YYY; in original form ZZZ}

\pubyear{2019}

\begin{document}
\label{firstpage}
\pagerange{\pageref{firstpage}--\pageref{lastpage}}
\maketitle

\begin{abstract}
Identifying the angular degrees $l$ of oscillation modes is essential for asteroseismology and depends on visual tagging before fitting power spectra in a so-called peakbagging analysis. In oscillating subgiants, radial (\lz) mode frequencies distributed linearly in frequency, while non-radial ($l\geq1$) modes are p-g mixed modes that having a complex distribution in frequency, which increased the difficulty of identifying $l$. In this study, we trained a 1D convolutional neural network to perform this task using smoothed oscillation spectra. By training simulation data and fine-tuning the pre-trained network, we achieved a 95 per cent accuracy on \kepler{} data. 
\end{abstract}




\section{Introduction}
Thanks to space missions such as \emph{CoRoT} \citep{baglin2006corot} and \emph{Kepler} \citep{borucki2010kepler}, a large number of precise photometric observations have been conducted for asteroseismic studies. Asteroseismology has been demonstrated as a powerful tool to investigate stellar interiors \citep[e.g.][]{chaplin2011ensemble,miglio2012galactic}. 

\begin{figure*}
	\includegraphics[width=\textwidth]{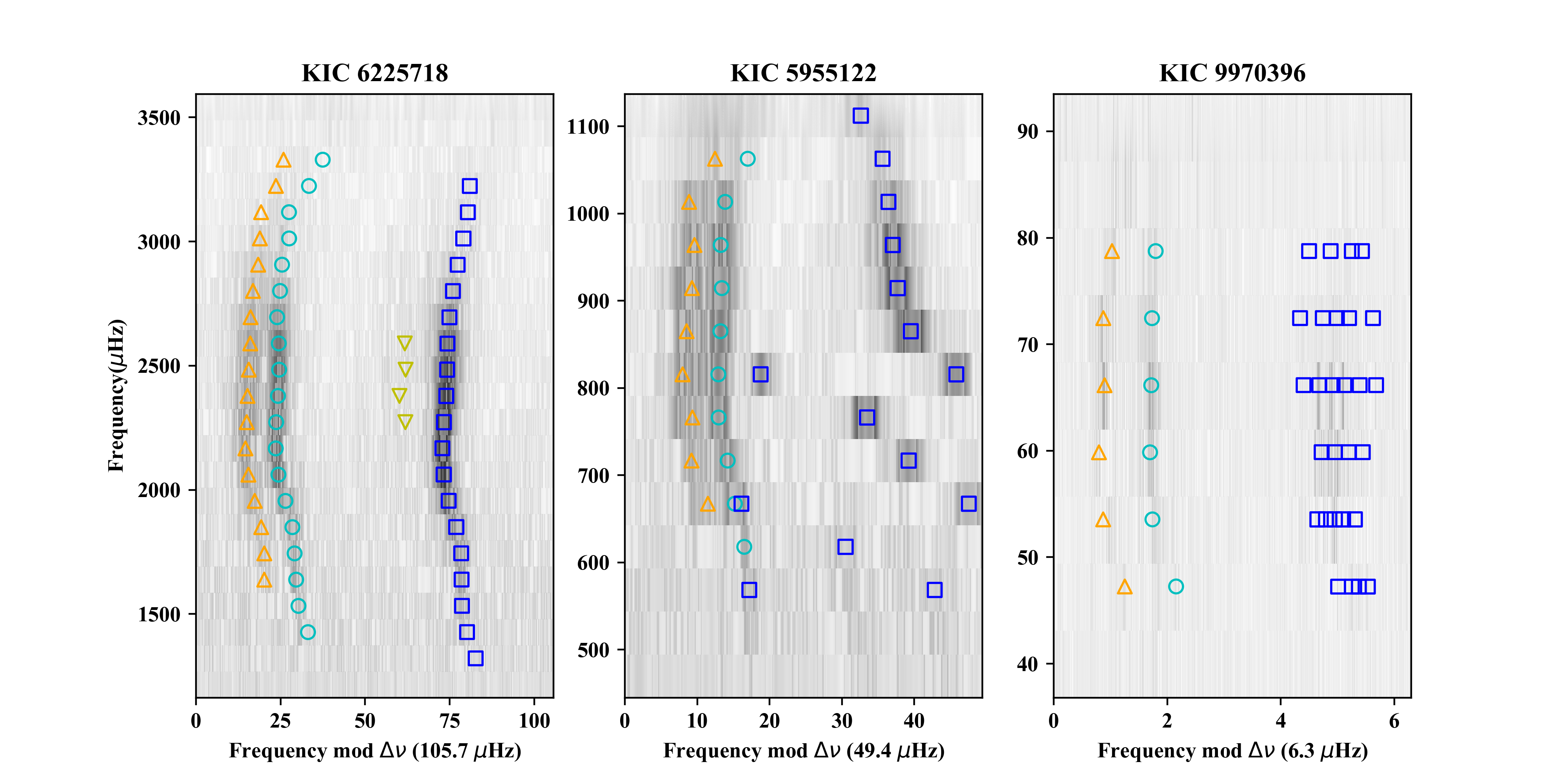}
	\caption{Typical \echelle{} diagrams of three stars including a main-sequence stars KIC 6225718 \citep[left panel,][]{aguirre2017standing}, a subgiant KIC 5955122 \citep[middle panel,][]{appourchaux2012oscillation} and a red-giant-branch star KIC 9970396 \citep[right panel,][]{li2017modelling}. The circles represent $l=0$ modes, the squares $l=1$, the upward triangles $l=2$, and the downward triangles $l=3$. The \lo{} modes in the subgiant (KIC $5955122$) shows a more complicated distribution than that in the main-sequence star and red-giant-branch star.}
    \label{fig:real_ech}
\end{figure*}

Solar-like oscillating main-sequence stars oscillate in pressure (p) modes \citep{Appourchaux2010}. According to the asymptotic theories \citep{tassoul1980asymptotic,gough1986ebk}, the p-mode frequencies of same $l$ are approximately equally spaced in frequency. The spacing, denoted by \dnu{}, is the so-called large separation. As central hydrogen exhausted, stars evolve into subgiants and later red giants, which show non-ridial ($l \geq 1$) oscillations of p-g mixed modes and radial oscillations ($l=0$) p modes. These mixed modes, coupled from g modes in the core and p modes in the envelope, do not follow regular spacings neither in frequency nor period and produce a more complicated power spectrum \citep{unno1979nonradial}. A common tool to identify mixed modes is to make \echelle{} diagrams, which slice the power spectrum into segments of equal length and stacked them upon each other. If the length of the segments equals to \dnu{}, the p modes will align nearly vertically. Such is not the case for the mixed modes. Fig.~\ref{fig:real_ech} shows typical \echelle{} diagrams of a main-sequence star, a subgiant and a red-giant-branch star observed by \kepler{}. In the typical ascending red giant branch stars and main-sequence stars, modes with same $l$-degrees have similar abscissa. Therefore, the mode identification assigning $l$ can take advantage of unsupervised cluster algorithms. However, modes in subgiants can not be automatically classified with such simplicity. The Bayesian multi-model fitting approach that models the stellar power spectral density with a mixture of Lorentzian profiles for as many oscillation modes as one intends to fit was widely applied to main-sequence stars and red-giant-branch stars. \citep[e.g.][]{benomar2009solar,kallinger2010asteroseismology,lund2017standing,vrard2018amplitude}. This method calculated the Bayesian factor and has only tested a small number of subgiants \citep[e.g.][]{benomar2012acoustic}. For accurate mode identification of subgiants, visual check based on the characteristics in the \echelle{} diagrams \citep{bedding2010scaled} is always necessary, but time costly. Thus, a more efficient algorithm is needed to deal with a large amount of data. The ensemble asteroseismology of red giants has been conducted for Kepler observations \citep{huber2010asteroseismology,mosser2012probing,vrard2016period,yu2018asteroseismology}. However, the studies for subgiants and main-sequence stars require an increase of the current sample size. With the arrival of space missions like TESS \citep{ricker2010transiting} and PLATO \citep{rauer2016plato}, more data for asteroseismic analysis would be available.

Individual frequencies are important parameters for asteroseismic analysis. They can constrain the stellar models \citep{aguirre2017standing} to accurately measure the fundamental stellar parameters,  for example, mass, radius and age \citep{chaplin2013asteroseismic,silva2015ages,ge2014asteroseismic}. The frequencies can also reveal the interior of the stars such as rotational profiles and convective mixing\citep{benomar2013properties,benomar2014asteroseismology}.  

End-to-end pipelines that extract the global oscillation parameters, mean large separation $\Delta\nu$ and frequency of maximum power $\nu_{\rm{max}}$ have been built and well developed \citep[e.g.][]{huber2009automated,mosser2009detecting,hekker2010octave,mathur2010determining,kallinger2010asteroseismology,garcia2014impact,stello2017k2}. In this study, we focus on building an algorithm to automatically perform mode identifications based on power spectra. Here, we choose a deep machine learning algorithm to due to its high accuracy, efficiency and great capability on processing high-dimensional data (i.e. frequency spectra). Deep learning methods have been proved useful on astronomy studies (e.g. light curve, frequency spectrum and stellar spectrum) including classifying evolutionary stage \citep{hon2017deep,hon2018deep}, detecting oscillations \citep{hon2018detecting} on frequency spectrum, determining stellar parameters on spectroscopic spectrum \citep{zhang2019938} and generating the solar magnetogram \citep{kim2019solar}.  

In this paper, we firstly construct labelled simulated frequency spectra with different $l$-degrees as images that contain spatial structure for deep convolutional neural networks to extract features for training a pre-trained model, and thereafter fine-tune the pre-trained model using half of the \kepler{} observed data to improve the performance of classifier. Finally, we use the rest of the \kepler{} observed data to test our model. We finally provide an effective and efficient classifier that identifies the $l$-degrees in subgiant stars. 

\section{Methods}
In this section, we describe how we built the training set and the test set for deep learning, followed by an introduction of deep convolutional neural networks. Then, we illustrate the detail structure of our classifiers used to assign $l$-degrees. We consider l=0,1,2 modes in this work. Note that l=3 modes are not considered because they constitute a small proportion of Kepler data.

Ideally, applying big observed data to train the model would be the best choice. \citet{li2020asteroseismology} calculated the power spectra of subgiants with good signal-to-noise ratios observed by \emph{Kepler} and identified about 1000 oscillation modes. However, the number of their sample is insufficient to train a deep learning model. Therefore, we used simulated data to achieve a pre-trained model and then used half part of the \kepler{} observed data to fine-tune the pre-trained model. The rest half part of observed data was used to carry out the test. 

\subsection{Constructing simulated power spectra}
\subsubsection{Oscillation frequencies} \label{section:echsimu}

We simulated oscillation mode frequencies using the p-mode asymptotic theory \citep{tassoul1980asymptotic,gough1986ebk} and a p-g mode coupling model proposed by \citet[hereafter D10]{Deheuvels2010}.
We first produced $l=0,1$ and $2$ p mode frequencies according to asymptotic relation\footnote{In subgiants, uncoupled \lz{} and 2 p modes are not observable. All $l\geq1$ modes have a mixed character. Here we refer the \lz{} and 2 p modes as the pure pressure modes assuming there is no coupling from the core.}. Then we produced the \lo{} p-g mixed mode frequencies using the coupling model. Note that l=2 modes in subgiants are also p-g mixed modes, but they have weaker coupling \citep{dupret2009theoretical}. Therefore we expect the frequencies of l=2 mixed modes still roughly follow the p mode asymptotic relation.

Firstly, we constructed $l=0, 1$ and $2$ p modes in nine radial orders around \numax{}. These modes follow the asymptotic relation \citep{mosser2012probing}:

\begin{equation}
	\nu_{n_{\rm p},l}=\Delta\nu(n_{\rm p}+\frac{l}{2}+\epsilon-d_{0l}+\frac{\alpha_{\rm curv}}{2}[n_{\rm p}-n_{\rm max}]^2)
	\\,
\end{equation}
where $n_{\rm p}$ is the p-mode radial order and $l$ is the angular degree. $\Delta\nu$ is the large separation and we here took \dnu{}=30$\mu$Hz as a fixed value for reasons that are illustrated in Section.~\ref{section:dataprepare}. The small separation $d_{0,l}$ is a function of \dnu. The item with $\alpha_{\rm curv}$ in the formula denotes the curvature pattern in \echelle{} diagram, in which the constant $\alpha_{\rm curv}$ represents the mean curvature of the p-mode oscillation pattern. $n_{\rm max}=\nu_{\rm max}/\Delta\nu$ and we here considered the fifth radial mode as $n_{\rm max}$. $\epsilon$ was treated as a fixed value because it does not affect the relative positions between modes.

The mean curvature $\alpha_{\rm curv}$ have a value of about 0.008 based on the red giants of \kepler{} observation \citep{mosser2012probing} and in our work, we used the following $C_{\rm curv}$ to recreate the curvature pattern,

\begin{equation}
	C_{\rm curv} = \frac{30}{11-10(\frac{n_{\rm p}-n_{\rm max}}{10})^2} \mu{\rm Hz} \\,
\end{equation}
where $n_{\rm p}=1$ to $9$ and $n_{\rm max}=5$. 

Thus, \lz{} modes frequencies with asymptotic relation are 
\begin{equation}
	\nu_{n_{\rm p},0} = n_{\rm p}+\epsilon+C_{\rm curv} \\.
\end{equation}
Then the frequencies of \lt{} modes are calculated using $d_{02}$, which is the mean distance between \lz{} ridge and its parallel \lt{} ridge. 
\begin{equation}
	\nu_{n_{\rm p},2} = n_{\rm p}+1+\epsilon-d_{02}+C_{\rm curv} \\.
\end{equation}
The frequencies \lo{} p modes conform to the relation of small separation $d_{01}$ of \lo{} modes from the midpoint of adjacent \lz{} modes. 
\begin{equation}
	\nu_{n_{\rm p},1} = (\nu_{n_{\rm p},0}+\nu_{n_{\rm p}+1,0})/2 + d_{01}(\nu_{n_{\rm p}+1,0}-\nu_{n_{\rm p},0})+C_{\rm curv} \\.
\end{equation}

Here, we used the result of \citet{bedding2010solar} that $d_{02}=0.125$ and set $d_{01}$ as a free parameter range from 0.01 to 0.04.

Then we produced the p-g mixed modes of \lo{} modes. D10 extent the analogy of p-g mixed modes proposed by \citet{Christensen-Dalsgaard97lecturenotes}. By writing the displacement of each oscillator as $y_i(t)=c_{i}\rm{exp}(-i\omega t)$ and solving an eigenvalue problem $AC=\omega^2 C$ with

\[
A=
\begin{bmatrix}
	\omega_1^2 & 0 & \dots & 0 &-\alpha \\
	0 & \omega_2^2 & 0 & \vdots  & -\alpha \\
	\vdots & & \ddots & & \vdots \\
	0& \dots & & \omega_{n-1}^2& -\alpha \\
	-\alpha & \dots & \dots & -\alpha & \omega_{n}^2
\end{bmatrix}
\],		
and $C=[c_1,\dots,c_n]$. We can calculate the eigenfrequencies of the system. In the matrix, $[\omega_1,\dots,\omega_{n-1}]$ represented p modes that we have calculate above. $\alpha$ is the so-called coupling term between the two oscillators. $\omega_n$ stood for a g mode. We assumed that the frequency of this g mode $\omega_n=\omega_{n-1}+c$ where $c$ is a free parameter that have four values, which represent that a g mode close to the previous frequency encountered with other p modes. Finally, we have 10 \lo{} p-g mixed  modes (9 p modes coupling with 1 g mode).

As stars evolve, multiple g modes can be coupled with p modes within the observable range. Here We only used one g mode to produce frequencies for simplicity. The total free parameters to construct the mode frequencies are \{$\alpha$, $\omega_n$, $d_{01}$\}. In particular, $\alpha$ was adopted from 0.2 to 0.5 with a step of 0.01. 

\begin{figure*}
	\centering
	\includegraphics[width=\textwidth]{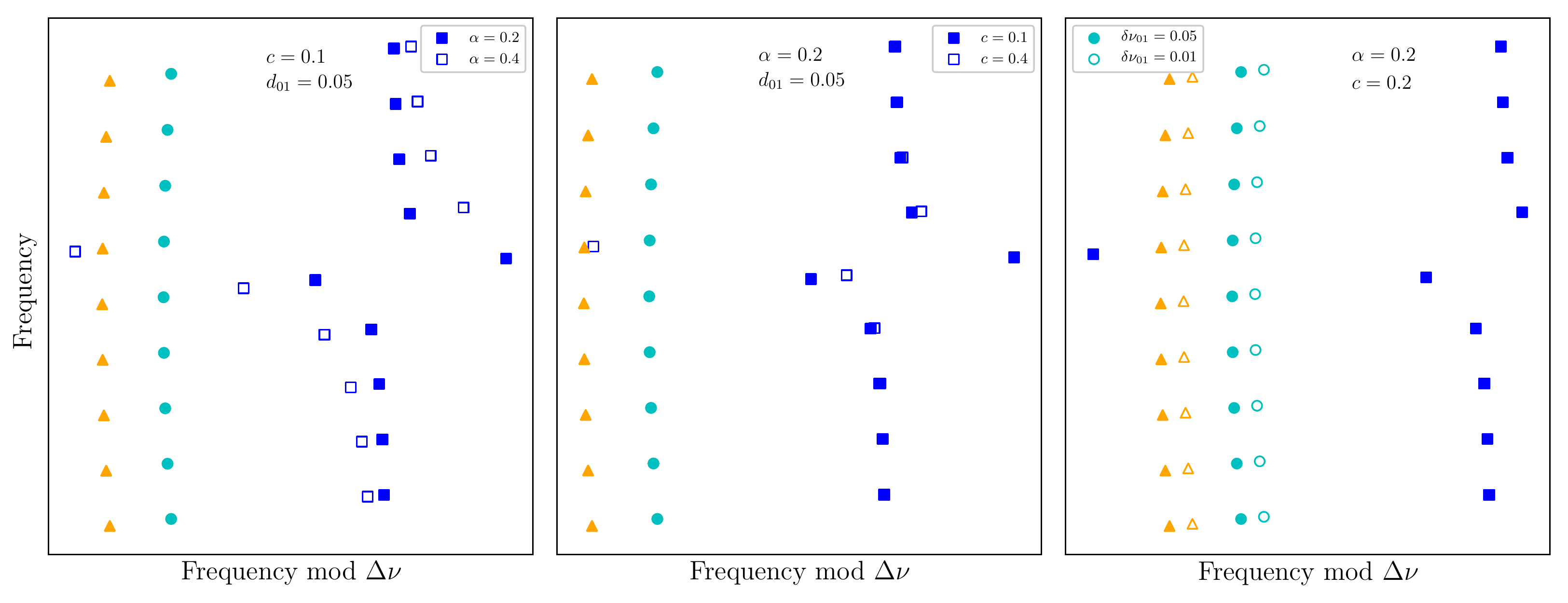}
	\caption{The \echelle{} diagrams with different free parameters, $\alpha$, $c$ and $d_{01}$. The three panels from left to right varies $\alpha$, $c$ and $d_{01}$, respectively with the two other parameters fixed. The symbols are the same as Fig~\ref{fig:real_ech}.}
    \label{fig:echelle}
\end{figure*}

Table~\ref{table:2} summarized the configuration of free parameters used to simulate mode frequencies. In total,  we produced $4\times4\times30=480$ power spectra that represented 480 subgiants. Fig.~\ref{fig:echelle} shows simulated frequencies on the \echelle{} diagrams with some configurations of $\alpha$, $c$ and $d_{01}$. For a set of parameter combinations, there are 9 \lz{} modes, 10 \lo{} modes, and 9 \lt{} modes, respectively, so there are 28 modes in total. Therefore, we achieved $480\times28=13440$ individual frequencies for a given effective temperature $T_{\rm{eff}}$ which will introduced in Section.~\ref{section:ps}. 

\begin{table}
	\centering
	\caption{Free parameters for simulating mode frequencies}
	\begin{tabular}{cc}
		\hline
		Parameter & Range \; |\;  Step \;|\; Sample number\\
		\hline
		$\alpha$   & $0.2\sim0.5 \; |\; 0.01\; |\; 30$ \\
		$c$        & $0.1\sim0.4$ \; |\; 0.1 \; |\; 4\\
		$d_{01}$ & $0.01\sim0.04 \; |\;  0.01\; |\; 4$\\
		\hline
	\end{tabular}
	\label{table:2}
\end{table}

In addition, using the frequencies of the stellar models is also a good choice and we have tested its performance. We chose the S4TESS catalog data \citep{ball2018synthetic} and models calculated by ourselves and they have showed the similar result. For the data of \citet{ball2018synthetic}, they calculated the frequencies using MESA \citep{paxton2010modules} and GYRE \citep{townsend2013gyre}. We selected 1222 subgiant stars based on effective temperature and $\Delta\nu$ and used the oscillation parameters they calculated based on stellar model. The result is not better than the method described above in this section. The main possible reason is the surface effect of the stellar models that effect the distribution of frequencies.

\subsubsection{Power spectra}\label{section:ps}
In reality, oscillation are detected at hundred $\mu$Hz to one thousand $\mu$Hz. For subgiant stars, the range of oscillation covers about a few hundred $\mu$Hz. From the sample of \citet{li2020asteroseismology}, we found a mean range of about 600 $\mu$Hz at about \numax=300$\mu$Hz. Thus, the parameters in this work were scaled to fit the frequency range.

We write the power spectrum as
\begin{equation}
	P(\nu)=L(\nu)\cdot w(\nu)
\end{equation} 
where $L(\nu)$ is a sum of Lorentzian profile that approximate the stochastic excitation of oscillation modes for well resolved modes. And $w(\nu)$ is a chi-square distribution with degree of freedom of two. The Lorentzian profiles can be described with three parameters height $H$, frequency centroid $\nu_0$ and line widths $\Gamma$
\begin{equation}
	L(\nu)=\frac{H}{(\frac{\nu-\nu_0}{\Gamma})^2+1}
\end{equation}
where frequency centroids were calculated as described in section~\ref{section:echsimu}. Because we would min-max normalize the power spectrum to build the training set, the absolute height $H$ of each mode is not essential. Therefore, we used the relative height equation for each order. \citet{salabert2011mode} measured the relative height of solar modes. However, \citet{benomar2013properties} found that relative heights in subgiants are different from the sun. The \lo{} relative heights have a larger range. The \lt{} relative heights are higher than in the sun. We here used the measurements from \citet{benomar2013properties}
\begin{equation} 
	H_{l=0}:H_{l=1}:H_{l=2}\approx1 : 1.6 : 0.65\label{heightt}
\end{equation}
The heights of oscillation modes on a same power spectrum are modified by a Gaussian profile centered at \numax{}, with a width of $\sigma$ and a height of $H_0$. The centroid $\nu_{\rm max}$ represents frequency of maximum power. Therefore, the height of modes nearby \numax{} is relatively higher than those far from \numax{}. Thus, we firstly assigned the height of \lz{} modes based on Gaussian profile, and then calculated the height of \lo{} and \lt{} modes using Equation~\ref{heightt}.

\begin{figure}
	\includegraphics[width=\columnwidth]{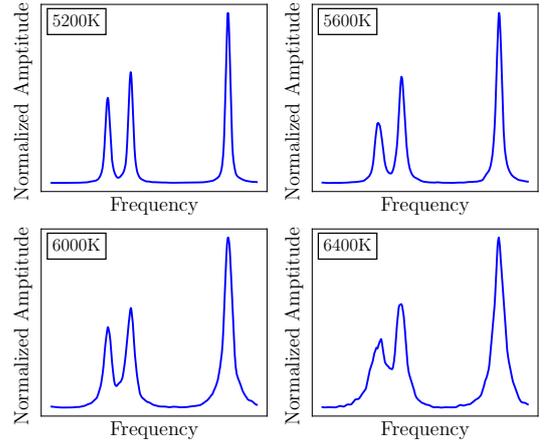}
	\caption{Simulated power spectra with four configurations of $T_{\rm eff}$, which changes the line widths.}
    \label{fig:ps4}
\end{figure}

The line widths $\Gamma$ at \numax{} are related to stellar effective temperature $T_{\rm eff}$ \citep{appourchaux2012oscillation}. We here used a 2 parameter relation derived using \citet{li2020asteroseismology} data for a subgiant sample.
\begin{equation}
	\Gamma_{\rm{max}}=a\times(\frac{T_{\rm eff}}{5777K})^b
\end{equation}
Although the line widths $\Gamma$ vary with frequency, we here didn't take into account because it failed to improve the accuracy of the model. We assigned $T_{\rm eff}$ with four values ranged from 5200K to 6400K in a step of 400K as shown in Fig~\ref{fig:ps4}. The total free parameters to simulate the power spectrum are listed in Table~\ref{table:3}. With four effective temperature we selected, the total number of simulated oscillation modes is $13440\times4=53760$.

\begin{table}
	\centering
	\caption{Free parameters for simulating the power spectrum. Gaussian distribution random number $\mathcal N$ is applied to produce tiny difference on the basis of conform to overall relations.}
	\begin{tabular}{cc}
		\hline
		Parameter & Range | Step\\
		\hline
		$T_{\rm eff}$   & $5200K\sim6400K$ | $400K$ \\
		$H_1$ & $\mathcal N(1.6,0.1)$\\
		$H_2$ & $\mathcal N(0.65,0.1)$\\
		$\Gamma_{\rm{max}}:a$ & $\mathcal N(1.47,0.17)$ \\
		$\Gamma_{\rm{max}}:b$ & $\mathcal N(8.18,4.96)$ \\
		\hline
	\end{tabular}
	\label{table:3}
\end{table}

\subsection{Data sets}\label{section:dataprepare}

We first smoothed the simulated power spectra and \kepler{} spectra using a 1$\mu$Hz wide boxcar (moving average) to demonstrate features more clearly. We found that the non-smoothed power spectrum increased the difficulty of training the convolutional neural networks. 

\begin{figure}
	\includegraphics[width=\columnwidth]{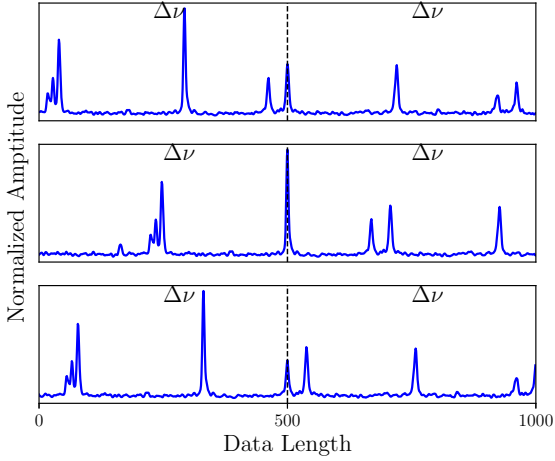}
	\caption{Segmented oscillation spectra for three samples. Each sample has 1000 points and covers $2$\dnu{} range. In the middle of array, the black dash lines indicate target modes to be classified which corresponds $l=0,1,2$ from top to bottom. }
\label{fig:modes}
\end{figure}

Then, we segmented the power spectra into 2-$\Delta\nu$-wide clips centred on the oscillation mode as shown in Fig.~\ref{fig:modes}, which compares spectral image representations between $l=0,1$ and $2$. Each sample is a $(1000,)$ 1-D array by linear interpolation. Since each sample is segmented into the same 2-$\Delta\nu$-wide 1D clips, we can fixed the $\Delta\nu=30\mu\rm{Hz}$ and changed other parameters ($\alpha$, c, $d_{01}$, $\Gamma$ and height ratio) to recreate characteristics of subgiant stars' power spectra because characteristic are scaled to the same wide range. The main purpose of simplifying the model is to reduce the difficulty of training.

The training set is composed of 53760 oscillation modes produced from simulated spectra and 500 oscillation modes from the \kepler{} sample. We used the former pre-trained model which can be considered as a base model upon which can be applied for optimization observed data from \kepler, $K2$ and \tess. This so-called fine-tuning technique \citep{tajbakhsh2016convolutional} was applied in our work because our simulated training subset can not replace real observation data. We fine-tuning the pre-trained model and produce the final model.

The remaining 569 individual frequencies made up the test set. For both the training set and test set, we min-max normalized the power spectrum of each star before segmenting due to the standard data processing in machine learning tasks.

Generally, a validation set is necessary for training the model and searching the most adaptive hyper-parameters. Here we simply randomly selected 10\% of the training set sample.

\subsection{Convolutional neural networks}

Artificial neural networks are common tools for solving for solving different kinds of problems especially for image processing and nature language processing. The artificial neural network is a mathematical representation inspired by biological neutron networks. Through mathematical algebra calculation, data flow in from the input layer and output the result from the last layer neurons. The input vector is $x=[x_0,x_1,\dots,x_d]$ which is the data we considered, and each neuron performed the calculation: $z=w^Tx+b$, where $w=[w_1,\dots,w_d]$ are weights and $b$ are biases, which are coefficients iteratively updated during the model training. 

The input vector is linear as described above. To solve non-linear problems, an activation function is needed. In this study, we used the \emph{Relu} activation function as a rectified linear unit, whose form is $f(x)=\max(0,x)$. This function is proved to learn the patterns and features efficiently in image processing tasks \citep{nair2010rectified}.

The convolutional neural network is a kind of feedforward neural networks with local connection, weight sharing and other characteristics which are great advantages compared with full connection neural networks. Local receptive field is one of the important ideas of convolutional neural networks, which is reflected in the form of \emph{convolution kernel} in mathematical form and is also called \emph{filter}. Convolutional operation of \emph{filter} and previous layer outputs the feature maps, which receives information within a small neighbored area of the previous layer and could extract elementary visual features. The filter slides in stride number and shares a single weight vector and a bias vector. In general, a convolutional layer always computes more than one feature maps with multiple filters and weight vectors. We write the output as:
\begin{equation}
	z^{({\rm lay})}=w^{({\rm lay})}\otimes a^{({\rm lay}-1)} + b ^{({\rm lay})}
\end{equation}
where ${\rm lay}$ means the layer number and $a$ is the convolutional kernel, whose size is a hyper-parameter. 

The objective of the convolutional neural network is to learn a particular set of weights from a training set that minimizes the error in approximating a ground truth y with a predicted output $\hat y$. Our study, a three class classification task with labels $0,1,2$. We used a normalized exponential function known as the \emph{softmax} function \citep{liu2016large} at the output layer to decide the class:
\begin{equation}
	p(y=c|x)=\frac{e^{w_c^Tx}}{\sum_{c=1}^C{e^{w_c^Tx}}}\,
\end{equation}
which defines the score of class $c$ out of $c = 3$ classes. The decision function is,
\begin{equation}
	\hat{y} = \rm{argmax}\ p(y=c|x)
\end{equation}
The \emph{softmax} function plays a role of the regression model considering it as a probability.

An error function is needed among machine learning models. For our classification task, we adopted the \emph{cross-entropy} \citep{shore1980axiomatic} to connect a true value $\hat{y}$ and a predicted value $y$,

\begin{equation}
	E(y,\hat{y}) = -\frac{1}{n}[y_i\log\hat{y_i}+(1-y_i)\log(1-\hat{y_i})]
\end{equation}
where $n$ corresponds to the sample number.
In order to minimize the error $E(y,\hat{y})$, a gradient descent algorithm was chosen. We used back-propagation (BP) algorithm \citep{hecht1992theory} defines error pass back to update the parameters, and other forms of artificial neural networks follow this approach \citep{lecun1995convolutional}. In our work, we select \emph{Adagrad} as the optimizer, an adaptive gradient algorithm from the improvement of stochastic gradient descent (SGD) algorithm \citep{duchi2011adaptive}. This optimizer speeds up the training and shortens the training time compared with SGD. After many iterations, the error function become smaller due to the effect of optimizer. Finally, the weight $w$ become stable and we obtained the classifier that could conduct classification tasks on test set and unclassified set.

\begin{figure}
\includegraphics[width=\columnwidth]{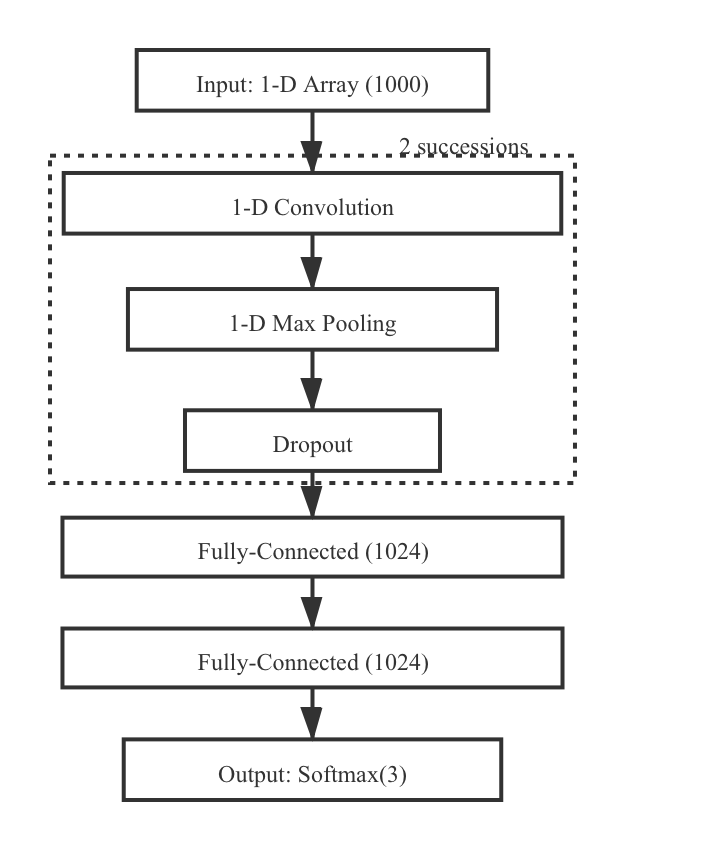}
	\caption{The construction of convolutional neural network illustrating the property of each layer and indicating the hyper-parameters applied to layers}
    \label{fig:model}
\end{figure}

The difficulty of building a useful neural network is to find a suitable structure and to find appropriate hyper-parameters, such as the size of filters, the number of filters and strides, etc. The architecture we applied is a classic convolutional neural network structure, which is similar to \emph{AlexNet} that contains \emph{convolutional} layers, \emph{maxpooling} layers and \emph{dropout} layers \citep{krizhevsky2012imagenet}. Among them, {\emph dropout} layers are to prevent overfitting problems by randomly manually set a percentage output to zero \citep{srivastava2014dropout} which is a hyper-parameter. We took this percentage value to 0.5, which showed good performance. Apart from these layers, we used full connection layers to combine the features extracted from the former layers together. The hyper-parameters optimization is always time consuming. We used the grid search approach to find the most suitable hyper parameters for our task. Fig.~\ref{fig:model} shows our classifier architecture.

We constructed the convolutional neural network with the deep learning framework Keras \citep{chollet2015keras} whose backend is Tensorflow \citep{abadi2016tensorflow}. Using Hyperas python module\footnote{https://github.com/maxpumperla/hyperas}, we found the optimal hyper-parameters. We used the filter size 64/16, filter number 16/16 and strides 3/1  for convolutional layers. The size of two fully-connected layers are both 1024. With a typical CPU in personal computer, predicting one thousand targets takes within one second.

\subsection{Fine-tuning the pre-trained network}

In order to make the classifier we modelled above more adaptive to the \kepler{}  observation data, we used half of the \kepler{} data to re-train and fine-tune the model. We first freeze the pattern extraction layers of two successions convolutional layers, max-pooling layers and dropout layers. And we just trained the fully connected layers whose role is to combine the patterns extracted using upper layers and find the most probable classification result. During the fine-tuning procedure, our base model rapidly converged and improved performance on real observed \kepler{} data, which reflected that the simulation data has the similar characteristics of observed data.

The fine-tuning model improved the accuracy of the base model to about 95\%.

\section{Results}
\label{sec:result}
\subsection{Classifier performance}

In our work, we trained two classification models. One was trained from all simulated data, which we call the pre-trained models. The other one is a fine-tuned model that combined the observation data with the pre-trained model. We here first used the confusion matrix in Table~\ref{table:cm1} and Table~\ref{table:cm2} to show the classification results of the two models separately.

\begin{table}
	\centering
	\caption{Confusion matrix of pre-trained model on the test set}
    \begin{tabular}{c|ccc}
    \diagbox{Actual}{Predicted} & \lz & \lo & \lt \\
    \hline
    \lz & 307                & 34                 & 6    \\
    \lo & 51                 & 406                & 10   \\
    \lt & 14                 & 19                 & 222  \\
	\end{tabular}
	\label{table:cm1}
\end{table}

\begin{table}
	\centering
	\caption{Confusion matrix of fine-tuning the pre-trained model on the test set}
    \begin{tabular}{c|ccc}
    \diagbox{Actual}{Predicted} & \lz & \lo & \lt \\
    \hline
    \lz & 172               & 6                 & 4    \\
    \lo & 8                 & 231               & 7   \\
    \lt & 1                  & 2                 & 138  \\
	\end{tabular}
	\label{table:cm2}
\end{table}

Based on the above confusion matrices, we have several indicators to evaluate our results. Each classification result is labelled with \emph{True Positive, False Positive, True Negative and False Negative}, which are commonly used in data science area. The definitions are:
\begin{enumerate}
\item True Positive (TP): The true class of a sample is $l$ and the model correctly predicts the class $l$..  

\item False Positive (FP): The true class of a sample is other, and the model incorrectly predicts it to be category $l$.

\item True Negative (TN): The true class of a sample is $l$, and the model incorrectly predicts it to be another class.

\item False Negative (FN): The real class of a sample is other classes, and the model is also predicted to be other classes. And we usually do not pay attention to this indicator
\end{enumerate}

We report several metrics calculated for the test set as follows:

Accuracy=$\rm{\frac{TP+TN}{TP+FP+TN+FN}}$

Precision=$\rm{\frac{TP}{TP+FP}}$

Recall=$\rm{\frac{TP}{TP+FN}}$

The accuracy shows the ratio of corrected classified sample and the total sample. The Precision is the fraction of relevant instances among the retrieved instances, while Recall (also known as sensitivity) is the fraction of the total amount of relevant instances that were actually retrieved. Table~\ref{table:1} shows the result.

\begin{table}
	\centering
	\caption{Matrix performance on the test set}
	\begin{tabular}{c|cccc}
		\hline
		Model & \multicolumn{2}{c}{Pre-trained model} & \multicolumn{2}{c}{Fine-tuned model}\\
		\hline
		Validation set accuracy & \multicolumn{2}{c}{98.9\%}& \multicolumn{2}{c}{92\%} \\
		\hline
		Test set accuracy & \multicolumn{2}{c}{87.5\%}& \multicolumn{2}{c}{95.1\%} \\
		\hline
		&Precision& Recall& Precision& Recall\\
		\lz&82.5\% &88.5\% &95.0\% & 94.5\%\\
		\lo&88.4\% &87.0\% &96.7\% & 93.9\%\\
		\lt&93.2\% &87.1\% &92.6\% & 97.9\%\\
		\hline
	\end{tabular}
	\label{table:1}
\end{table}

\subsection{Performance analysis}
\begin{figure}
\includegraphics[width=\columnwidth]{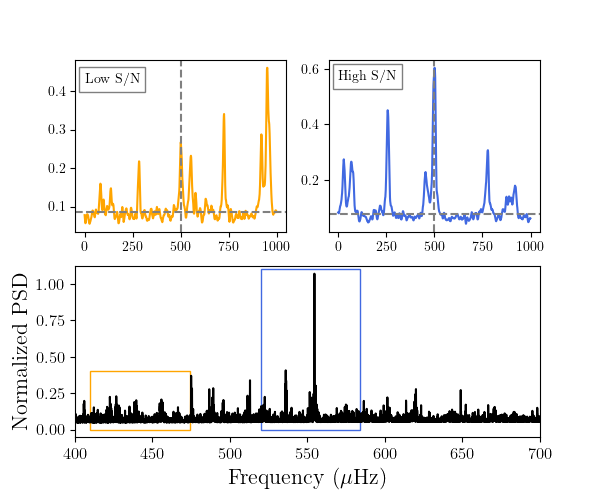}
	\caption{The bottom panel shows the normalized power spectra of KIC 11771760 which is one of our test set sample. The left orange rectangular covers the lower S/N modes and the right blue rectangular covers the higher S/N modes that corresponds two samples in our test set. The upper two panels demonstrate the range of this two rectangular. The horizontal grey dashed line represents the median of the array and the vertical dashed grey line indicates the target mode. Dividing the target mode height by the median of the sample to get the S/N ratio.}
    \label{fig:defsnr}
\end{figure}

\begin{figure}
\includegraphics[width=\columnwidth]{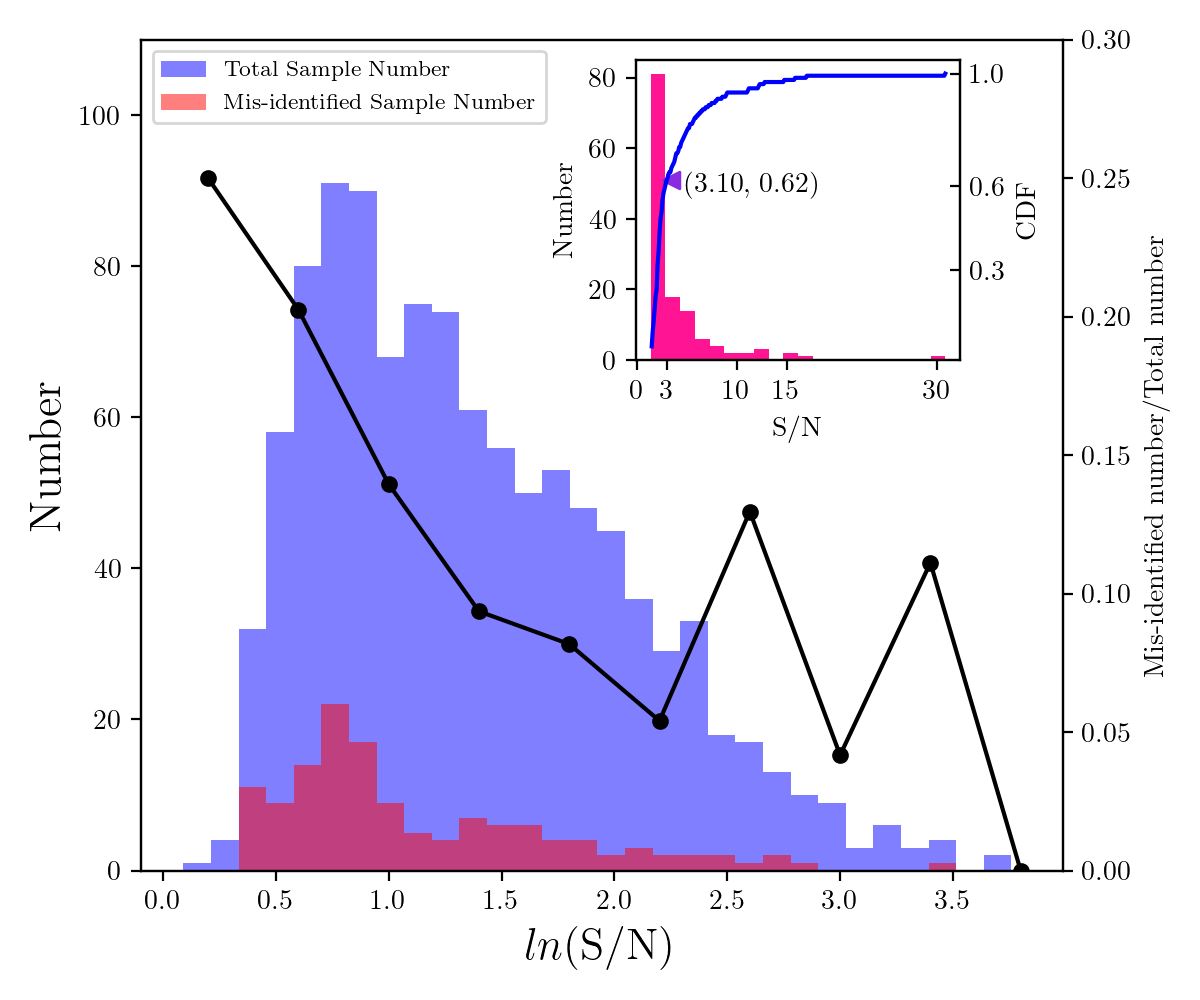}
	\caption{The S/N distribution of \kepler{} sample tested on pre-trained model. The inset figure shows the distribution of mis-identified sample. The blue solid line represents the cumulative distribution function and left violet triangle indicated the point of 60\% volume that corresponds to a region below S/N=3}
    \label{fig:snr}
\end{figure}

\begin{figure}
\includegraphics[width=\columnwidth]{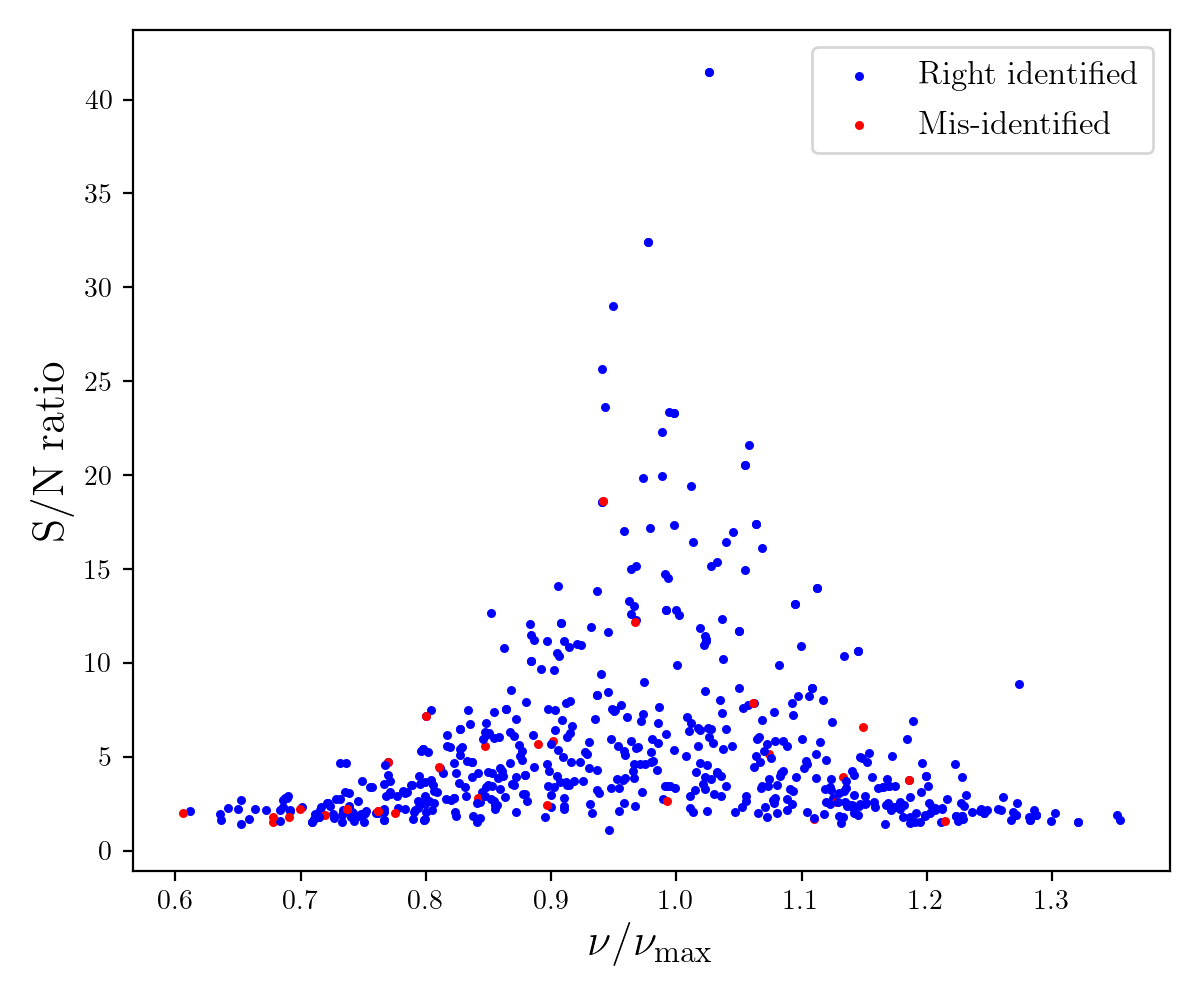}
	\caption{The S/N distribution of modes vs the distance to \numax.}
    \label{fig:snrdis}
\end{figure}

The performance of the base model is not perfect on \kepler{} data. The likely reason is that the simulation data are a little different from observable data. Firstly, simulation data have high signal-to-noise ratio (S/N), whereas the S/N of \kepler{} data covered a large range as shown in Fig.~\ref{fig:snr}. However, adding low S/N ratio samples to the training samples will greatly increase the difficulty of training the model. The definition of S/N here is a local value that divided the height of target mode by the median of the (1000,) array as showed in Fig.~\ref{fig:defsnr}. Fig~\ref{fig:snr} shows that most mis-classified samples are low S/N. The accuracy for sample S/N<3 and S/N>3 is 81\% and 90.5\%, respectively. Secondly, we took approximation during simulation. For example, in our simulation, the relations of relative height between different $l$ degree modes are very strong prior assumption. However, 69\% low S/N false predicted samples are the modes less than 0.8\numax{} as shown in Fig~\ref{fig:snrdis}. These modes are affected by the background contributed to granulation and facula and white noise, therefore they become obscuring for machine to extract features. The difference like this between the simulation data and observed data is reduced by fine-tuning the pre-trained network that improves the classifier's performance. 

At the same time, the performance on the test set is expected to differ from the training set because the two have different parameter distributions. The fine-tuning model can reduce the difference between these two sets and improved the performance on classifying low S/N modes. However, we should still take care of the results of low S/N modes because most of the mis-classified samples still concentrated in this area.

\begin{figure*}
	\includegraphics[width=\textwidth]{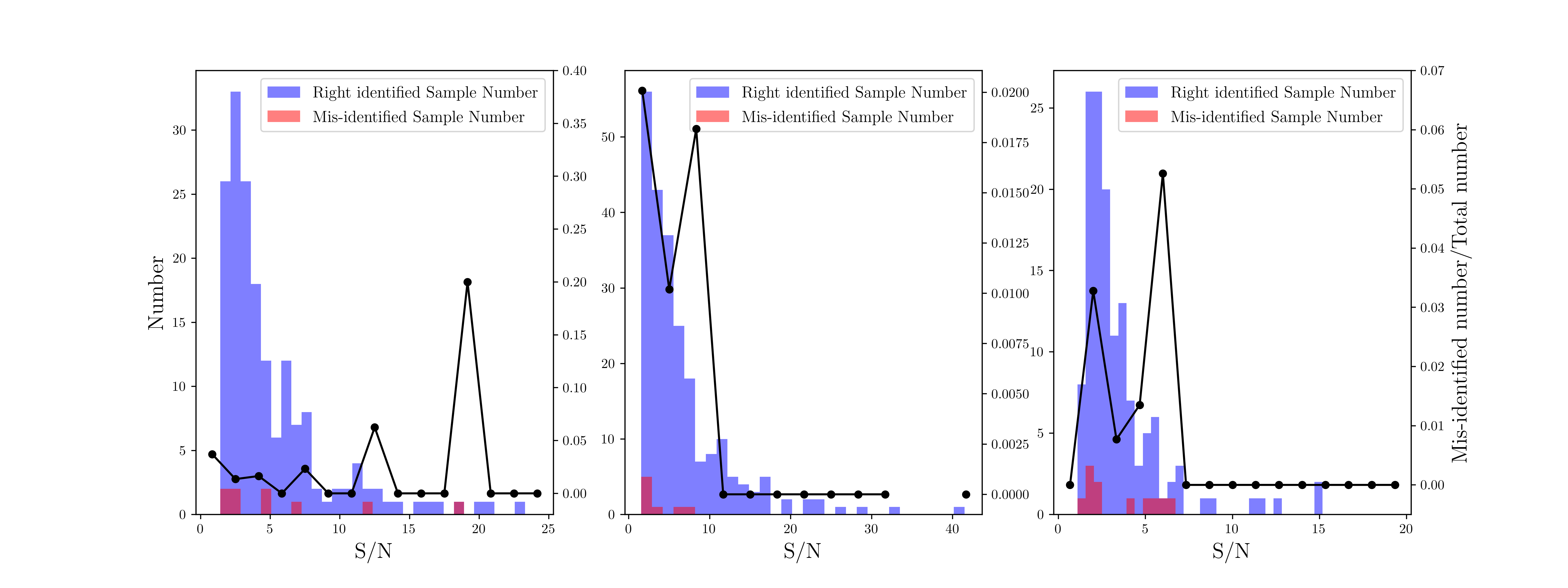}
	\caption{The S/N distributions of modes predicted as $l=0,1,2$ (from left to right) tested using the fine-tuned model. }
    \label{fig:8}
\end{figure*}

\begin{figure}
\includegraphics[width=\columnwidth]{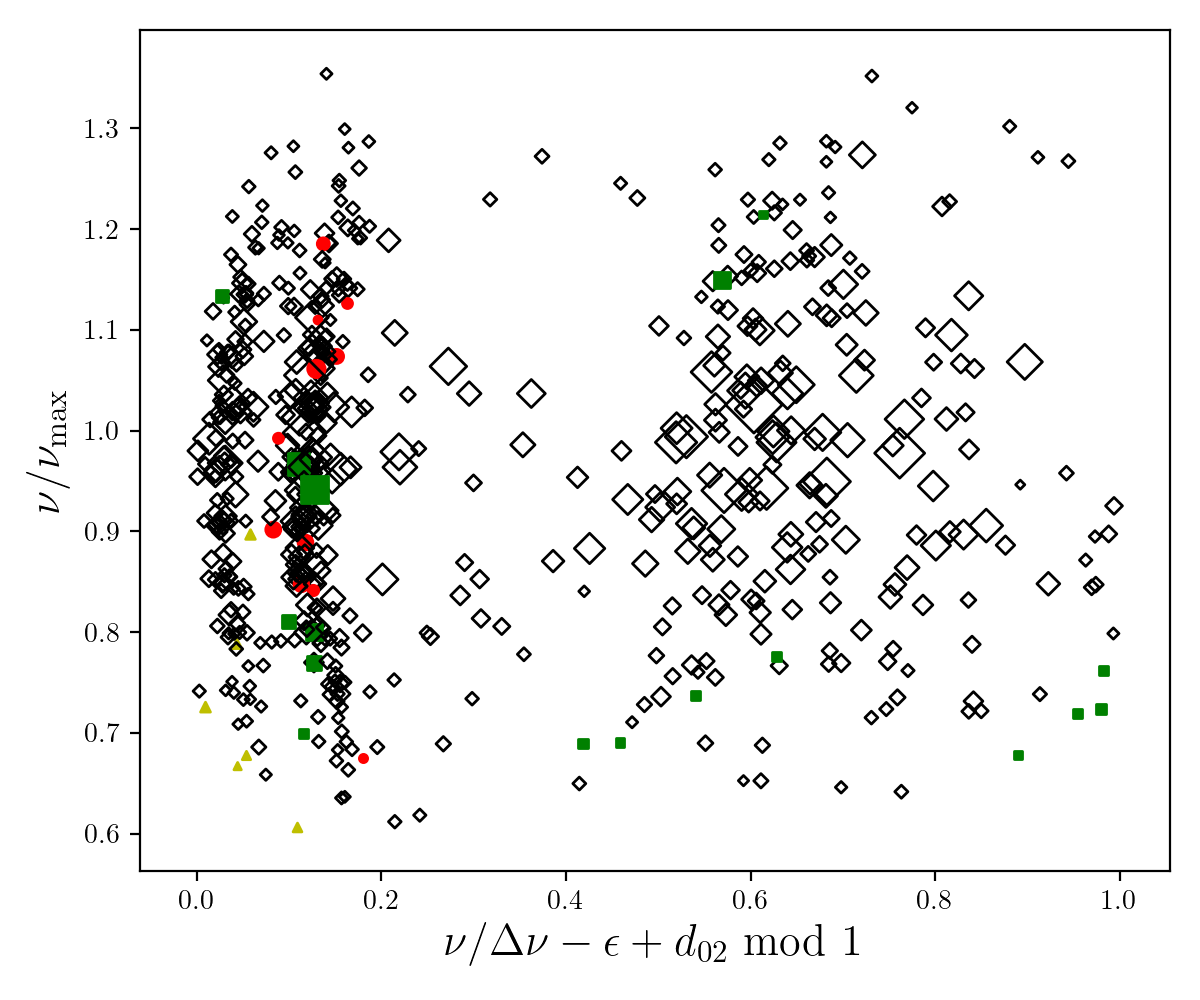}
	\caption{The classification result of the fine-tuned model plotted on the scaled \echelle{} diagram. The symbols indicate mode in the test set, with symbol size proportional to S/N. The filled symbols are mis-classified modes, with the red circles showing $l=0$ modes, green squares $l=1$ and yellow triangles $l=2$.}
    \label{fig:norm_ech}
\end{figure}

The fine-tuned model have a 95 per cent accuracy. \lo{} modes have a relative high precision 96.7\% than \lz{} and \lt{} modes, which means that modes predicted as \lo{} are most credible. On the contrary, \lt{} modes have a higher recall compared with \lo{} modes and the lowest precision, that is to say, they are not easy to be predicted wrong, but other modes are easily predicted as \lt. Fig.~\ref{fig:8} and Fig.~\ref{fig:norm_ech} show the distribution of true predicted and false predicted samples with corresponding S/N. The $l=0,1$ mis-classified modes focus at the lower S/N range. Only a small number of high S/N \lo{} modes are mis-classified because they are very closed to \lz{} modes within 1$\mu$Hz and always predicted as \lt{} modes. The corresponding \lz{} modes are correctly identified. However, this leads to an increased probability that the \lz{} mode in other locations is identified wrong. In conclusion, the classification of $l=0,1$ modes with good S/N of a threshold can be trusted. And the predicted \lt{} modes should be carefully treated. Unfortunately, there is no way to warn of possible mistakes. 

\subsection{Usage of models}
In order to make use of future \tess{} and \emph{PLATO} data and to identify the angular degree of subgiant stars to conduct further ensemble asteroseismology, we suggest the following procedure:
\begin{enumerate}
    \item Obtain the power spectra, mode frequencies and global oscillation parameters (\dnu{} and \numax{}) of subgiant stars.
    \item Prepare the data as illustrated in section~\ref{section:dataprepare}.
    \item Use fine-tuned model to identify the angular degree of modes. 
\end{enumerate}

Considering that data will be released gradually, we can use the fine-tuned model with initially. When the data is big enough, we can also train a new model entirely based on observation data. We estimated that three to four thousand accurate labelled sample is enough for a reliable model according to test we conducted on \kepler{} data which means we need an additional observation of about 100 subgiants.


\section{Conclusion}

We have applied a convolutional neural network to perform the identification of subgiants' oscillation modes. We used smoothed segmented power spectra as samples, whose patterns are learned by the convolutional neural networks model. Test on \kepler{} data demonstrated a 99\% cross-validation accuracy and an 87\% accuracy.  By fine-tuning the model, the accuracy on test set improves to 95\%. Note that it is difficult to perform accurate classification on low signal to noise ratio modes. With future bigger data, more accurate labelled data are available and the model could be update or trained entirely with observation data.

Most of samples in our test set are early type subgiant stars, therefore, the effect of the model we trained has been tested on early-type subgiant stars. Its performance on late type subgiant stars needs more samples to test. \citet{appourchaux2020attempting} proposed a new method to automate the identification of mixed dipole modes for subgiant stars and their method work well on late type subgiant stars.

The method in this work can also provide reference for the mode identification of red giant branch stars and red clump stars. The biggest challenge for classify the modes of red giants is rotation splitting. The current model ignored the rotation effect to the power spectrum. However, it’s normal in red giants. In the future, a 2D model perform on \echelle diagram maybe a good choice.

\section*{Acknowledgements}

This work is supported by the Joint Research Fund in Astronomy (U2031203 and U1631236)
  under cooperative agreement between the National Natural Science 
  Foundation of China (NSFC) and Chinese Academy of Sciences (CAS). This work has also received funding from the European Research Council (ERC) under the European Union’s Horizon 2020 research and innovation programme (CartographY GA. 804752). We would also like to thank Jie, Yu and group at The Beijing Normal University for discussions.

\section{Data availability}
Data available on request.



\bibliographystyle{mnras}
\bibliography{mybib.bib} 







\bsp	
\label{lastpage}
\end{document}